\documentstyle[11pt,aaspp4]{article}
\tighten
\lefthead{Jun}
\righthead{Radio Emission from Young Supernova Remnants}

\begin{document}

\def\etal{{\it et al.~}}
\def\eg{{\it e.g.,~}}
\def\ie{{\it i.e.,~}}
\def\gsim{\raise0.3ex\hbox{$>$}\kern-0.75em{\lower0.65ex\hbox{$\sim$}}}
\def\lsim{\raise0.3ex\hbox{$<$}\kern-0.75em{\lower0.65ex\hbox{$\sim$}}}


\title{Radio Emission from a Young Supernova Remnant Interacting with an
Interstellar Cloud: MHD Simulation with Relativistic Electrons\altaffilmark{3}} 

\author{Byung-Il Jun\altaffilmark{1} and T.W. Jones\altaffilmark{2}}
\affil{Department of Astronomy, University of Minnesota, \\
    116 Church Street, S.E., Minneapolis, MN 55455 \\
    and                                   \\
University of Minnesota Supercomputer Institute, \\
1200 Washington Avenue South, Minneapolis, MN 55415}

\altaffiltext{1}{current address : Lawrence Livermore National
Laboratory, P.O. Box 808, L-630, Livermore, CA 94551, email : jun2@llnl.gov}
\altaffiltext{2}{email : twj@msi.umn.edu}
\altaffiltext{3}{Accepted for publication in the Astrophysical Journal}

\begin{abstract}
We present two-dimensional MHD simulations of the evolution of a young
Type Ia supernova remnant during its interaction with an interstellar
cloud of comparable size at impact. We include for the first time
in such simulations explicit relativistic electron transport. This
was done using a 
simplified treatment of the diffusion-advection equation,
thus allowing us to model injection and acceleration of cosmic-ray
electrons at shocks and their subsequent transport. From this information
we also model radio synchrotron emission, including spectral information. 
The simulations were carried out in spherical coordinates with azimuthal 
symmetry and compare three different situations, all incorporating an
initially uniform interstellar magnetic field oriented in the
polar direction on the grid. In particular, we 
modeled the SNR-cloud interactions for a spherical cloud on
the polar axis, a toroidal
cloud whose axis is aligned with the polar axis and, for comparison,
a uniform medium with no cloud.

We find that the evolution of the overrun cloud qualitatively resembles that
seen in simulations of simpler, but analogous situations; that is, the
cloud is crushed and begins to be disrupted by Rayleigh-Taylor and
Kelvin-Helmholtz instabilities. However, we
demonstrate here that, in addition, the internal structure of the SNR 
is severely distorted as such clouds are engulfed. That  has
important dynamical and observational implications. The principal
new conclusions we draw from these experiments are: 
1) Independent of the cloud interaction, the SNR reverse
shock can be an efficient site for particle acceleration in a young
SNR.
2) The internal flows of
the SNR become highly turbulent once it encounters a large cloud.
3) An initially uniform
magnetic field is preferentially amplified along the magnetic
equator of the SNR, primarily due to biased amplification  in
that region by Rayleigh-Taylor instabilities. A similar bias
produces much greater enhancement to the magnetic energy in the SNR
during encounter with a cloud when the interstellar magnetic field
is partially transverse to the expansion of the SNR.
The enhanced magnetic fields have a significant radial component,
independent of the field orientation external to the SNR.
This leads to a strong equatorial bias in synchrotron brightness
that could easily mask any enhancements to electron acceleration 
efficiency near the magnetic equator of the SNR. Thus, to establish
the latter effect it will be essential to establish that the magnetic
field in the brightest regions are actually tangential to the blast wave.
4) The filamentary
radio structures correlate well with ``turbulence-enhanced'' magnetic
structures, while the diffuse radio emission more closely follows the
gas density distribution within the SNR. 
5) At these early times the synchrotron spectral index due
to electrons accelerated  at the primary shocks should
be close to 0.5 unless those shocks are modified by cosmic-ray
proton pressures. While that result is predictable,
we find that this simple result can be significantly
complicated in practice by SNR interactions with clouds. Those
events can produce
regions with significantly steeper spectra. Especially if
there are multiple cloud encounters, that can lead to
nonuniform spatial spectral distributions, or, through
turbulent mixing, produce a spectrum difficult to relate to
the actual strength of the blast wave.
6) Interaction  with the cloud
enhances the nonthermal electron population in the SNR in our simulations because
of additional electron injection taking place in the shocks associated
with the cloud.
Added to the point number (3), this means that SNR-cloud encounters
can significantly increase the radio emission from the SNR.
\end{abstract}

\section{Introduction}

Shell supernova remnants (SNRs) have been traditionally imagined in
simple terms as spherical explosions within a homogeneous
interstellar medium. To explain their diverse characteristics
the classic cartoon (Woltjer 1972) divides the
dynamical evolution of SNRs into four distinct phases dependent on
the ratio of the swept-up ISM mass to the ejected mass and the
degree to which the expansion remained adiabatic.  In this picture
young remnants would consist of ``freely expanding'' ejected stellar 
debris surrounded by a 
skin of shocked ISM, while older remnants would consist largely of the
swept-up material concentrated  by geometrical factors and  eventually
even more by radiative losses into a thin shell immediately inside a blast wave. 
Ultimately, the
expansion would slow to sonic velocities, allowing random motions
within the ISM to disperse the remnant. Today the above simple construct
is recognized as wholly inadequate to explain the rich structures, 
kinematics and dynamics observed in real SNRs.
The boundary between ejecta and swept -up material
is often Rayleigh-Taylor (R-T) unstable, so that the two become mixed (\cite{gull73}).
In addition it is now clear that the transitions between Woltjer's 
distinct phases are
often likely to be at least as long lived as the stages themselves (\eg
\cite{cmb88}; \cite{d-pj96}).
It is not obvious that the distinct phases actually develop very often, in fact.
This is especially the case for the transition out of the ``free expansion''
phase to an adiabatic blast wave. 
The above features are made even more significant by the fact that
most, if not all, SNRs
exist within highly inhomogeneous media, and that their evolution and
appearance are strongly influenced by this feature virtually from
the moment they are born (\cite{chu94}).
This realization was a dominant theme during a recent workshop on SNR
dynamics in Minneapolis (\cite{snrmeet}).

In this context the need is clear for major progress in our
understanding of SNR dynamics, especially during stages when both
stellar ejecta and swept-up material influence dynamics.
The problems are made more complex by their inherent multidimensionality, 
especially, but not exclusively, when the external medium is inhomogeneous. 
Fortunately, modern computational tools are now being applied with some
success to this challenge. For example, Tenorio-Tagle \etal 1991
have carried out 2D hydrodynamical (HD) simulations of young Type II 
SNRs interacting with (spherical) wind-blown bubbles. \cite{bork97}
have used 2D HD computations to study the anticipated collision between the 
ejecta of SN 1987A and its circumstellar ring. Similarly,
\cite{d-pj96} explored the dynamics of a young Type II SNR
developing on the edge of a giant molecular cloud. Jun (1995) (see 
also \cite{jjn96}) carried out fully MHD 2D simulations of
a young Type Ia SNR expanding into an ISM containing many small
clouds. These studies all confirm that motions and structures  
within the SNRs are made much more complex by the presence of
the external inhomogeneities. 

To connect dynamical theory with actual observations it is necessary to
model emissions that can be detected at earth, of course. 
Again, modern computational tools
are now facilitating that step. This is especially true for
models of thermal line and continuum emissions. 
\cite{bork97} included such 
predictions in their 2D simulations of 1987A, for
example, and  there is a substantial literature of sophisticated
thermal SNR emission models based on 1D dynamical simulations (\eg
\cite{bork96} ). 
On the other hand progress has been relatively slow to
model nonthermal emissions from SNRs, even though the
most detailed observational information available often comes from 
nonthermal synchrotron emission, particularly
at radio frequencies (\eg \cite{weil88}; \cite{snrmeet}). Synchrotron models for SNRs are largely 
limited by our abilities to compute meaningful magnetic field
structures and to model the acceleration and
transport of the
relativistic electrons responsible for the emission. Magnetic field
properties, including strength, in young SNRs are almost
certain to be dominated by multi-dimensional dynamical effects, especially
instabilities (\eg \cite{jun96a}; \cite{jun96b}). Even accepting as
fact the most 
popular diffusive shock acceleration (DSA) model for production of 
relativistic electrons in SNRs the task of calculating the
electron distribution in space and energy within a SNR is also not trivial.
Consider just the simplest version of DSA
theory in which the particle energy distribution is a power-law
whose index is determined entirely by the density jump through
the accelerating shock. In a realistic young SNR model there are two or
more shocks present (a forward and a reverse shock, plus secondary
shocks generated by inhomogeneities), and the strengths of those
shocks are likely to vary in both space and time. Nonspherical
internal motions transporting the inhomogeneous electron population 
will add to the complications. Any prediction of synchrotron
emission must include multidimensional treatments of
both the particles and the fields, and how they interact.
That point was clearly evident in our earlier 2D 
simulations of DSA associated with plane-shocked interstellar clouds 
(\cite{jon93}) and supersonic gas clouds (\cite{jkt94}).
However, until now no multidimensional SNR  computations have been
published that include explicit treatment of electron acceleration.

The present paper is an effort to begin to fill the gaps
outlined above. We have carried out fully MHD simulations in 2D of
young Type Ia SNRs within an inhomogeneous medium. Since 
computations already exist to illustrate what happens during the
interaction of an SNR with clouds that are both very small (\eg \cite{kmc94},
\cite{jjn96}) and very large (\eg \cite{af93}; \cite{d-pj96})
compared to the SNR, we deal here for the first time directly with the 
intermediate case, when
the cloud and SNR are of a similar size. To explore the influence of the
cloud geometry we consider both spherical clouds and clouds
with a ``ring''  or toroidal geometry. As a first meaningful step towards
the inclusion of full relativistic particle transport we introduce
a new simplified scheme for electron transport, including the
effects of DSA. Despite its simplicity, the scheme offers a powerful
tool for exploring the dynamics in many complex flows.
In \S 2 we describe our electron transport scheme, while the 
remaining numerical
details are outlined in \S 3. The computational results are 
discussed in \S 4 and
a short conclusion is presented in the final section.

\section{Simplified Method for Electron Transport}

Since no comparable treatments of relativistic electron transport
have appeared in the literature, we briefly outline a simple
but powerful scheme for following electron transport that should be useful in
the SNR context.
Let us assume that nonthermal electrons are accelerated 
by the DSA process mentioned above. While DSA is not yet firmly
established as the only important SNR electron acceleration process, it is hard to imagine
that it does not take place in the strong SNRs shocks, and
recent nonthermal X-ray observations in several SNRs seem to support its
presence (\eg \cite{koy95}; \cite{reynolds96}; \cite{allen97}). 
The key to DSA is rapid angular scattering of charged
particles 
on MHD waves, keeping energetic particle distributions almost
isotropic. If their scattering lengths are long compared to
those of thermal particles, however, these particles will
diffuse spatially if they are not uniform.
The applicable transport equation can be written as (\eg Skilling 1975)
\begin{equation}
{df \over dt } = {1 \over 3} p {\partial f \over \partial p}(\nabla
 \cdot \vec v) + \nabla \cdot (\kappa \nabla f) + Q,
\label{dceq}
\end{equation}
where $f(\vec x, p, t)$ is the isotropic part of the
electron distribution, $\kappa$ is
the spatial diffusion coefficient, $Q$ is a source term that
represents the net effects of ``injection'',  ``escape''
and radiative losses at a given momentum,
while $d~/dt$ is the Lagrangian time derivative. 
The particle momentum is $p$, while the thermal plasma velocity is
$\vec v$.
The first term on the right accounts
for adiabatic compression or rarefaction in the background flow.
Equation [\ref{dceq}] must be solved simultaneously with the MHD
equations for the bulk flow in order to follow nonthermal electron
transport and determine the synchrotron emissivity. 
Since equation [\ref{dceq}] is a function of particle
momentum as well as space and time, it can add considerably to
the computational difficulty and expense of a problem, especially because the diffusion
coefficient for the electrons is likely to be a function of
momentum. The added momentum dimension is particularly important in multidimensional
flow treatments, since they are often already computationally
demanding.
However, for many astrophysical applications, the particle
energy distributions should be very broad. We can take advantage of 
that feature to introduce a 
simplified treatment of equation [\ref{dceq}].
The central point is that in many cases the nonthermal electron spectrum is 
a power-law to a good approximation over at least moderate-sized momentum bins. 
If for $p_i \le p \le p_{i+1}$, we assume
$f(p) = f_i~ (p/p_i)^{-q_i}$, then
we can integrate equation [\ref{dceq}] within each such momentum bin to
obtain
\begin{equation}
{dn_i \over dt } = - {1 \over 3} (\nabla \cdot v ) q_i n_i + \nabla
\cdot (<\kappa> \nabla n) + Q^\prime
\label{dndt}
\end{equation}
where $n_i = 4 \pi \int_{p_i}^{p_{i+1}} p^2 f dp$ is the number of
electrons in the bin, $<\kappa> =
{\int p^2 \kappa \nabla f dp \over \int p^2 \nabla f dp}$, and $Q^\prime
= 4\pi  \int_{p_i}^{p_{i+1}} p^2~Q dp$. 
In principle the evolution of the particle distribution could be
followed in arbitrary detail with equation [\ref{dndt}], like equation [\ref{dceq}],
using a sufficiently fine grid of
momentum bins, but its real advantage lies in the fact that $q_i$
often varies slowly with momentum, so that only a few broad bins may 
capture the basic structure of $f(p)$.
Note if we apply equation [\ref{dndt}] on a fixed momentum range that 
the adiabatic compression term (first on the r.h.s.) provides implicitly for the flux of
particles into or out of that momentum range under the assumption that
the slope of the distribution is continuous at the computed momentum boundaries.

From the above definitions we have
\begin{equation}
n_i = 4 \pi {{f_i p_i^3} \over {q_i - 3}}
[1~-~({{p_i} \over {p_{i+1}}})^{q_i-3}].
\label{nieq}
\end{equation}
Since $f(p)$ is continuous we can also write $f_{i+1} = f_i (p_i/p_{i+1})^{q_i}$.
Once $n_i$ is updated from an initial distribution using equation [\ref{dndt}],
then equation [\ref{nieq}] can be solved to update  $q_i$ exactly if we know $f_i$.
That procedure becomes straightforward if the two lowest momentum
bins have the same value of $q$ and have the same width in
log of momentum space; that is if $q_1 = q_2$, and $p_3/p_2 = p_2/p_1$.
Then
\begin{equation}
q_1 = q_2 = 3 - {{\ln{(n_2/n_1)}}\over{\ln{(p_2/p_1)}}}.
\label{q1q2}
\end{equation}
With this information $f_1$, $f_2$, $f_3$ and hence, recursively, all 
remaining values of $f_i$ and $q_i$ can be found. 
If $q_{i+1}\ne q_i$ for $i>2$ an iterative solution to equation [\ref{nieq}]
can be obtained for each $i>2$ using equation [\ref{q1q2}] as an initial guess.
Even if $q_1 \ne q_2$
we have applied a variant of this method successfully to obtain approximate
updates for $q_i$, assuming that $q_i$ varies smoothly through
successive bins.
In principle the method we have outlined can be applied to both 
electronic and ionic cosmic-ray populations,
even when the population provides a dynamically significant backreaction
pressure. For the present simulations, however, we will treat only
electrons and consider them to be a passive population, without
significant pressure.
If we assume a pure power-law electron
distribution (that is, $q_i = q$) exists at each location it is somewhat more
convenient to extend the momentum bound $p_3 \to \infty$.
Then we replace equation [\ref{q1q2}] with
\begin{equation}
q = q_1 = q_2 = 3~+~{{\ln{(1~+~n_1/n_2)}}\over{\ln{(p_2/p_1)}}}.
\label{qinfty}
\end{equation}
That is the version of this scheme we apply below.
Keep in mind, of course, that $q$ may vary with time and location.

Electrons important to centimeter radio emission
in SNRs have energies below about
10 GeV, assuming the magnetic field is more than 
about 10 $\mu$Gauss. Since there is at most small
spectral curvature seen in the local intensity in 
SNR radio emission (\eg \cite{rey92}) it is a reasonable first
approximation to assume that $f(p)$ is a single power law at
a given location and time within the SNR. In that case we need to track
only two momentum bins.
For such low energy electrons there is another suitable
approximation that greatly reduces the effort required for electron
transport. In particular, the scattering lengths of these
electrons should be very small compared to the size of cells within any
practical numerical grid, and the  time to accelerate them in
shocks very much shorter than time steps (helping explain
why there is little spectral curvature, in fact). 
The Larmour radius, $r_L$, for
a 10 GeV electron, which should be comparable to the relevant
scattering lengths in a strongly turbulent plasma (\cite{jonell91}), 
$r_L \sim 10^{13}/B_\mu$cm, where $B_\mu$ is the
magnetic field strength in $\mu$Gauss. 
The time to accelerate this particle through DSA in a shock
of velocity, $u_s$, is roughly (\cite{lag83}) $t_a \sim r_L c/u_s^2 \sim 1/(B_\mu~u_{s3}^2)$ years,
where $u_{s3}$ is expressed in units of $10^3~$km/sec.
Under these conditions, we can ignore the diffusive term in equation
[\ref{dceq}] away from shocks and can assume that the electron
distribution within a given fluid element achieves its equilibrium 
DSA form as it passes through a shock. 
The form will change subsequently only if a stronger shock is encountered.
Thus, electrons in the postshock flow have a power law slope
$q_1 = q_2 = min(3r/(r - 1), q_{init})$, where $r$ is the compression ratio through
the shock  and $q_{init}$ is the upstream slope of the distribution (\eg \cite{drury83}). 
Since our simulations involve adiabatic gas flows with $\gamma = \frac{5}{3}$,
then $r \le 4$ and $q \ge 4.$
With these assumptions our transport method becomes very simple and
computationally inexpensive. A conceptually similar scheme using a less
accurate method for updating $q$ in continuous parts of the flows was 
applied previously 
by Jones \& Kang (1993) and by Jones \etal (1994) (see also
\cite{and94}) to multidimensional
simulations of DSA around plane-shocked clouds and
supersonic gas clumps, respectively.
Those simulations also included a dynamically consistent two-fluid 
treatment of the cosmic-ray protons, but we have omitted that feature 
from our present set of calculations.
Our new method is potentially much more broadly applicable
than the earlier one, however. For example, a variant of it has
been used recently to study electron transport in radio
galaxies (Jones, Ryu \& Engle 1998). There with only 8 momentum
bins it was possible to simulate electron spectra subject to
strong synchrotron cooling, or ``aging''. Those authors also discuss
in much more detail the
basis for this scheme and present some tests.

We note here a comparison with the earlier electron transport
scheme introduced by Jones \& Kang (1993). That scheme used a
crude method for following the spectral informtion; namely, 
advection of a weighted spectral index. The current method is
more physically based and correctly uses the direct properties
of the distribution itself. Thus, it is more accurate. Further, 
the new method can be used even when the electron spectra are
not true power laws, whereas the previous scheme could not.

Cosmic-ray ions are thought to be injected directly as a natural
part of the physics of collisionless shocks (\eg \cite{jonell91}). 
Although it is
more difficult for electrons to be injected in this way, it seems
likely that a small fraction of the thermal electron flux
through strong shocks will also be injected (\eg \cite{lev96}). 
To model this we use the 
same simple injection scheme adopted, for example, by \cite{kj91} and \cite{jon93}.
A small, fixed fraction, $\epsilon$, of the thermal electron flux through
the shock is converted into a power law with the appropriate
equilibrium slope for that shock.
In particular we assume (see eq. [\ref{dndt}])
\begin{equation}
Q^\prime = \sum_i Q_i^\prime = {{\epsilon W_s}\over{\mu_e m_H}},
\end{equation}
where $W_s$ is the Lagrangian shock velocity, $\mu_e$ is the electron
mean molecular weight, $m_H$ is the mass of the proton. Note that the
ratio $Q_{i+1}^\prime/ Q_i^\prime$ equals $n_{i+1}/n_i$
in equation [\ref{nieq}].

\section{MHD Numerical Methods and Initial Conditions}

The gasdynamical and magnetic field variables were  
followed through the ideal MHD equations for an adiabatic $\gamma = \frac{5}{3}$ plasma,
solved using the ZEUS-3D code developed at the National Center for
Supercomputing Application (Clarke \& Norman, 1994).  The algorithms of
this code are described in detail by Stone \& Norman (1992a,b).  In
brief, ZEUS-3D is a fully three-dimensional Eulerian MHD code.   Shocks are
stabilized by von Neumann-Richtmyer artificial viscosity.  Fluid is
advected through the mesh using the upwind, monotonic interpolation
scheme of van Leer (1977).  Magnetic fields are transported
using the constrained transport method (Evans \& Hawley 1988), modified
with the method of characteristics (Stone \& Norman 1992b).  The code
is further modified and improved by us for the
moving grid method and stable magnetic field transport.  The stable
algorithm for the magnetic field evolution developed by Hawley \&
Stone (1995) is implemented in order to fix a known problem for
weak magnetic fields and demonstrated to perform well (see Clarke 1996 for a
discussion of the weak field problem).
The advection equation for electron number density is solved by
the same algorithm as the one used for the fluid.  The adiabatic compression
term in equation [\ref{dndt}] is updated implicitly for the time centered 
number density of electrons in each momentum bin.  
For electron acceleration we detect only shocks with Mach number, $M\ge2$.

The computation is carried out in a two-dimensional spherical polar
coordinate system. 
The computational plane ($r-\theta$) is resolved uniformly with either 
300x300 cells or 600x600 cells.   
Except for the presence of the large interstellar
cloud and the geometry of the magnetic field the situation modeled is
the same as that explored by \cite{jun96a}.
We have also included DSA of electrons, of course, which they did not.
A Type Ia supernova
explosion is initialized with kinetic energy $10^{51}$ ergs, where
the ejecta velocity is assumed to be linearly proportional to the radius.
The inner $\frac{4}{7}$ of the ejected stellar mass ( total 1.4 $M_{\odot}$) is set up
with a constant density and the outer $\frac{3}{7}$ of the stellar mass is
initialized with a power-law density profile $\rho \propto r^{-7}$.
The background ISM is initialized with a uniform density $\rho_i = 1.67 \times
10^{-24} gm^{-3}$ and temperature $T_i = 10^4 K$, giving a
pressure $p_g = 2.24 \times 10^{-12}$ dyne/cm$^2$. There is a uniform 
magnetic field, $B_i = 3.54 \mu G$, in the
direction of the symmetry axis. Thus, the field lines are parallel to the
SNR expansion at the pole (along the symmetry axis) and 
perpendicular at the equator.
The ratio of gas to magnetic pressures in the ISM is $\beta = \frac{p_g}{p_b} \approx 4.5$.
The outer radius of the supernova at the start of our simulation
is $0.1 pc$.  The entire
computational plane is perturbed with a random density noise of $2.0 \%$
amplitude.  
A moving grid method is used to follow the expanding SNR.
This allows us to resolve the
initial profile of supernova material accurately and to follow the 
evolution over a large
dynamical range of SNR expansion.  In order to
introduce an interstellar cloud in the background, we update the
outer boundary in the r-direction every time-step.   Therefore, the cloud
comes in from the outer {\it r} boundary as the supernova shock expands.  We
place the cloud center at $r_{cc}=2.0 pc$. Its radius on the computational
plane is $r_{co} = 0.5 pc$. The cloud is assumed to be in pressure
equilibrium with the ISM, to have a uniform density, $\rho_c = 5\rho_i$
and to be threaded uniformly by the background magnetic field.
With these initial conditions the SNR blast wave first encounters the
cloud at a time, $t \approx 100$ years.
Three different cases are simulated.
First, as a control we computed the evolution of this
SNR in a uniform medium, without a cloud present. In the second
case we placed the cloud center on the symmetry
axis of the grid ($\theta = 0$), so that it has a spherical form.
In the third case the cloud center is centered at a polar angle
$\theta = {\pi \over
4}$.  This last cloud is actually a ring  or torus when projected in the 
azimuthal direction around the symmetry axis.
We evolve a passive, Lagrangian invariant 
quantity (mass fraction) to distinguish ISM and stellar material and
to follow their contact discontinuity.
The cloud material is
also followed by a separate mass fraction function so that we can see
the cloud disruption clearly.

The nonthermal electron distribution followed during these
simulations is divided into two
momentum (or energy) bins, as outlined in the previous section.  
The first momentum bin is bounded by $p_1 \approx  m_ec$ and $p_2 \approx 2 \times 10^3 m_ec$
(1 MeV $\le E \le $ 1 GeV), whereas the second bin extends the electron
energy $E$ to infinity.  We include both ``pre-existing''
electrons assumed to be
present in the ISM as part of the galactic cosmic-ray population
and electrons injected at shocks within the SNR, as discussed above.
The pre-existing electron number density 
above 1 MeV is assumed to be $10^{-8} cm^{-3}$, roughly 
consistent with estimates for the ISM from $\gamma-$ray observations
(\eg Moskalenko \& Strong 1998).
Although the ISM spectrum does not appear to be a simple power law,
its actual form has little importance here, since that population is 
converted into a power law spectrum with $q \approx 4$ once it passes through
the SNR blast wave at the early times we are modeling.
Thus, for convenience we assume for the pre-existing electrons a 
power law ($f \sim p^{-q}$) with $q = 4.7$.
The shock injection rate at $keV$ energies, $\epsilon$, was taken to be
$\epsilon = 10^{-4}$. While electron injection efficiency at shocks
is very uncertain, we shall point out below that the injected population in our
simulations for any rate $\epsilon > 10^{-5}$
would dominate the pre-existing population. Thus, since the electrons
are passive, the main role of $\epsilon$ is as a scale factor for
the synchrotron emission from the SNR, as discussed in \S 4.3.3. Typical estimates for the
analogous injection rate for CR protons are in the range $\epsilon_p \sim 10^{-3}~-~10^{-2}$,
(\eg \cite{ber96}). Then, $\epsilon/\epsilon_p \sim 10^{-2}$,
consistent with the ratio of these two components in the galactic CR
population.
The $\epsilon = 10^{-4}$ injection rate translates into an
injection at strong shocks of $10^{-7}$ of the thermal electron flux into 
the $E > MeV$ bin.

\section{Results}

\subsection{One-dimensional Evolution of Relativistic Electrons}

The production and evolution of the relativistic electron
populations is qualitatively similar in all the models we ran, including the
case without an interacting cloud. One-dimensional dynamics of
the latter is easier to follow and also to see the relative roles of
pre-existing and injected electrons, as well as the relative
roles of the two primary shocks in the SNR simulations.
Figs.1a-e show for such a one-dimensional calculation the evolution of the
relativistic electrons.
The gas density evolution is plotted in Fig.1f for comparison. 
Readers are referred to \cite{jun96a} for a discussion of
the dynamics of similar SNRs and for plots of the evolution of
other physical variables such as gas pressure.
The simulation shown here, as well as our two-dimensional ones were carried
to a time, $t = 500$ years, when the swept-up mass from the
uniform ISM is about 4 times
the ejected stellar mass. Dynamically the SNR is still a long ways
from possessing a self-similar, Sedov structure (see, \eg Fabian \etal 1983;
\cite{d-pj96}), but by this time most of the ejected
mass has passed through the reverse shock, and the outer blast wave 
motion approximates an $r\propto t^{\frac{2}{5}}$ law.
Fig.1a shows for our simulation the total electron density, $n_{e1}$, 
including contributions from both shock-injected and pre-existing populations 
in the $1 MeV \le E \le 1 GeV$ bin.
The radii of the reverse shock, contact
discontinuity, and supernova forward shock are designated by R.S.,
C.D., and F.S., respectively. 
The density, $n_2$, in the higher energy bin, $E> 1GeV$ is shown in Fig. 1c.
The behaviors of $n_1$ and $n_2$ are virtually identical, because the spectral
index for the electrons is nearly a constant $4.0$ between R.S. and F.S, as
seen in Fig. 1e.
The electron density interior to the reverse shock, R.S.,
is very nearly zero, since the original population has been extremely
rarefied by expansion of the ejecta.
Curiously, then, the electron  density just outside (downstream of) the reverse
shock is much higher than inside the forward shock at early
stages (see the electron number density profile at t=100 years).   This
feature results from faster injection of thermal electrons through the
reverse shock than through the forward shock.  There are two reasons for this.
First, the
upstream gas density of the reverse shock is higher than for the
forward shock as seen in the first (t=100 years) plot of Fig.1f.
Second, the inward velocity of the reverse shock moving through the expanding
supernova material is higher than the velocity of the forward shock
through the ISM.
Together these cause the flux of electrons eligible for DSA to be
higher at the reverse shock.
Incidentally, the Mach number of the reverse shock is also higher than the Mach
number of the forward shock, although for both shocks the spectral index of
emerging electrons is very close to the limiting value, $q = 4$
throughout the simulation.  Actually, the reverse shock Mach number
increases during this stage of SNR evolution, since the interior gas 
is strongly cooled by adiabatic expansion.
The importance of the higher reverse shock speed is well illustrated
at t =300 years when $n_e$ is higher at the
reverse shock than the forward shock even though the gas density upstream
of the reverse shock is lower than for
the forward shock. During these early stages of the SNR evolution,
therefore, the
reverse shock is more efficient in accelerating (mainly injecting) the
particles.  While $n_e$ continues to be greatest behind the
reverse shock to the end of our simulation the contrast
diminished with time because it becomes rarefied through
expansion.
This result is very interesting, since the possible importance of
the reverse shock as a source of high energy particles as been mostly
ignored. At least one previous calculation has seen supporting
behaviors. Close examination of Dorfi's (1990) one-dimensional 
simulations
of SNR evolution including a two-fluid treatment of CR protons
clearly reveals a substantial population of CRs produced at the
reverse shock during similar stages of evolution. His diffusion coefficient,
however, was large, so that CRs produced at the two shocks almost
blended together. Nevertheless, the CR pressure exhibits a distinct
peak at the reverse shock.

An obvious issue is the relative importance of electron populations
inside the SNR coming through local shock injection and from the
pre-existing galactic population. For the parameters we chose
this can be evaluated in Fig. 1.
We ran a comparison simulation with $\epsilon = 0$, so that only
pre-existing electrons are present inside the SNR.
The case without injection is shown in Figs.1b and 1d.
Also for comparison
the dotted lines represents the number density without the
the effects of DSA at the shock.  Therefore, the electron number
densities at the shock are increased exactly by the shock
compression ratio and decreased downstream of the shock due
to the adiabatic expansion.  The evolution of electrons including the
acceleration process (the redistribution of particles by
flattening the power law of the particle distribution) is shown with
the solid lines.  In our simulation, the electron momentum power law index
changes from 4.7 to 4.0  at both shocks (see Fig.1e) as the
particles are processed by DSA.  For this case, in fact, the spectral index
is always 4.0 between the two shocks and 4.7 elsewhere. This
flattening of the power law transfers electrons from the
lower energy bin ($1~MeV < E < 1~GeV$) to the second bin ($E> 1~GeV$). 
That effect primarily accounts for the difference between the dotted line and solid line in
Fig.1d.    From equation \ref{nieq} we can see if we assume the total
number of electrons remains fixed that the 
redistribution of the
electrons in the higher momentum bin can be expressed as 
$n_2(d)/n_2(u) \approx (p_2/p_1)^{q_u-q_d}$,
where $u$ and $d$ refer respectively to the upstream and downstream
values. In this case that is roughly a factor of 130.
There is a secondary difference between the solid and dotted curves in
Fig. 1b and Fig. 1d coming from the adiabatic expansion of the flow
between the shocks.
That influences the particle distribution through the adiabatic term $- {1 \over 3}(\nabla \cdot v)q n$
in equation \ref{nieq}.
This term reflects the difference between particle fluxes into and out
of a fixed momentum bin as individual electrons change their momentum
at a rate $\dot p = - \frac{1}{3} p \nabla \cdot v$.
The fluxes are more closely balanced when the momentum distribution
is flat; \ie, when $q$ is small, so adiabatic effects are reduced in 
the post-shock regions.
Conversely, the adiabatic effects are very large interior to the
reverse shock, since $q$ is large there. This accounts for the
much more dramatic reduction in that region for both $n_1$ and $n_2$ 
when compared to the gas density.

\subsection{Dynamics of a Shocked Cloud}

   The interaction between a steady, plane shock wave and a dense cloud has been studied
in great detail by a number of authors (\eg Woodward, 1976; \cite{mckee75}; 
Stone and Norman 1992; Jones and Kang 1993; Klein et al. 1994; Mac Low et
al. 1994; Xu and Stone 1995).  We briefly review their findings as
background for our new results involving the more complicated situation
of a young SNR over-running a cloud of comparable size.
As a shock sweeps past the cloud, the incident wave is
reflected back into the shocked background material, creating a bow shock.
The incident shock is also transmitted into the cloud forming a
so-called ``cloud shock'', thus, flattening
the cloud into a pancake shape.  
The cloud shock propagates with
velocity, $v_c \simeq ({\rho_i \over \rho_c})^{\frac{1}{2}} v_s$ where, $\rho_i$,
$\rho_c$, and $v_s$ are the density of the intercloud medium, the
density of the cloud, and the shock velocity in the intercloud medium,
respectively.    The slower velocity of the cloud shock relative to the
intercloud shock results in a shear layer at the cloud surface, where
the Kelvin-Helmholtz instability subsequently develops.  On the other
hand, the Richtmyer-Meshkov instability occurs on the front surface,
due to the impulsive acceleration of the cloud front by the shock. 
Although these instabilities cause erosion of the cloud, wholesale
destruction of it is delayed until after the cloud shock exits at
the cloud rear.
A rarefaction wave is then
reflected back into the cloud leading to its re-expansion,
while the whole cloud is effectively accelerated by
the pressure difference between the front and back of the cloud.
That acceleration leads to development of the Rayleigh-Taylor (R-T)
instability on the front of the cloud.
Dense R-T fingers form from cloud material and rarefied cavities
begin penetrating the cloud body.
The combined effects of these instabilities will lead, in the absence
of a substantial magnetic field, to destruction of the cloud on a
timescale several times greater than the so-called cloud crushing time,
$t_{cc} = ({{\rho_c} \over {\rho_i}})^{\frac{1}{2}}{r_c \over v_s} \approx
{r_c \over v_c}$.
To a first approximation this description also applies to our
simulations of a cloud-SNR encounter.
For that interaction
the cloud nominal crushing time is approximately 230 years, compared to an
cloud-SNR encounter interval during our simulations of about 400 years.

There is one additional feature within the cloud that has some importance
to our discussion of synchrotron emission. In addition to the very strong
shock that enters the cloud directly as it impacts with the SNR shock,
another, somewhat weaker shock penetrates into the cloud from its
perimeter. This shock results as the external shock wraps around
the cloud, increasing the pressure on its sides. By the Bernoulli
effect and the fact that the flow around the cloud is almost sonic, the
lateral pressure on the cloud is considerably less than the pressure on its ``nose''.
However, as long as the primary shock is strong, there should be
sufficient pressure to drive a weaker shock into the cloud from this side.
Although this shock has not received much attention in the literature
of shocked clouds, it is clearly present in published images (\eg Fig. 1 of \cite{jon93}).
The importance of this shock to our later discussion comes from the
fact that material processed through this shock is also
subsequently stripped from the cloud by instabilities and mixed
with material ISM material that was processed even earlier by the SNR
blast wave.

\subsection{New Two-dimensional SNR Simulations}

\subsubsection{Hydrodynamical Evolution}

As mentioned before, we have simulated two different SNR-cloud 
encounters, involving spherical and ring cloud geometries, in addition to
a comparison case including only a uniform ISM.  
At the start we note that the hydrodynamical evolution is
similar for both spherical and 
ring clouds. However, the behavior of the magnetic field is very 
different in the two cases, as we discuss later.  
The hydrodynamical similarity mirrors results of Klein et al. (1994) 
and Xu \& Stone (1995)  who compared
the hydrodynamical evolution of spherical clouds to cylindrical clouds
with their axes parallel to the incident shock normal. Those authors
found the cloud to be somewhat more flattened in 
the cylindrical case, but behavior patterns were not otherwise 
sensitive to the initial shape of the cloud.
We focus our discussion on the ring cloud, 
which minimizes the artificial axial-focusing that can result from
the axisymmetric assumption. We will provide appropriate 
comparison to the other cases, as needed.

Before the interaction with the cloud occurs, the
SNR shows the typical double-shock structure 
mentioned in discussing the one-dimensional simulation.
As noted by many before us, the contact discontinuity 
separating the stellar ejecta and the swept-up ISM is R-T
unstable. From that, R-T fingers develop on the shell of ejecta, extending 
outward into the shocked ISM.
In our simulations, the blast wave hits the cloud when $t \approx 100$ years.
The ensuing evolution of the interaction is summarized in Fig.2.  
Density, cloud mass fraction, and magnetic field strength are
shown from left to right.  Epochs shown from top to bottom are
t=200, 300, 400, and 500 years.  
The cloud evolution is similar to that seen in simulations
of plane-shock cloud encounters, although it is made somewhat more complex
by the follow-up collision of the cloud with the shell of stellar ejecta. 
The separation between the SNR shock and the ejecta shell is smaller
than the size of the cloud at the time of impact.
Thus, on a timescale much less than $t_{cc}$ following the first
cloud encounter, the 
cloud bow shock runs into the ejecta, quickly compressing it
and sweeping portions of it in front of the cloud.
The SNR reverse shock then merges with the cloud bow shock, so
that it ``stands'' in front of the cloud as the rest of the
expanding remnant wraps around.
Portions of the ejecta shell and, most visibly, distorted R-T
fingers are dragged into the flow around the cloud.
The cloud shock is strong, compressing the cloud by about a factor of four.
On the other hand, the bow shock is weak, with a 
compression ratio in the shocked ISM material 
$\sim 1.5~-~1.6$ in our simulation.
It is straightforward to show, in fact, that the compression in 
shock-induced bow waves is always less than about 2  for
adiabatic $\gamma = \frac{5}{3}$ shocks (\cite{jon93}).
By the end of our simulation the cloud begins to show signs of disruption
due to the K-H, Richtmyer-Meshkov and R-T instabilities that
we mentioned in the previous section. The distortions of the
cloud boundary are quite evident in the mass-fraction images of
Fig. 2. Note, as a result of the instabilities, that the sharp
boundary between cloud and non-cloud material evident at $t = 200$ years
has begun to show clear indications of mixing in the later mass fraction images.
Based on the previous work cited above, we expect that after
a few more hundreds of years the cloud would become highly disrupted.

Viewed from the perspective of the evolution of the SNR we see that
the encounter with the cloud severely distorts the expansion of the
blast wave, but more importantly the expansion of the stellar 
ejecta and the reverse shock. 
From the known behavior of plane shock-cloud encounters we 
expect that the blast wave itself will eventually recover close to spherical
form unless there are multiple encounters (on the latter see, 
for example, \cite{jjn96}).
The cloud should fragment as it is engulfed by the SNR, and become
mixed with ejecta, subject to
magnetic field constraints on these processes (\eg \cite{mmksn94}; \cite{jrt96}).
On the other hand, the cloud disruption generates vorticity within the SNR and
that should lead to large and long-term aspherical motions. Especially
if there are multiple cloud encounters, these could have major impact
on the appearance and kinematical structure of the SNR (again, see \cite{jjn96}
for the analogous situation for an SNR sweeping multiple, small clouds).
Thus, this study supports and amplifies our previous conclusions that unless
young SNRs are evolving within truly uniform environments, they
will likely exhibit substantial nonspherical structures, especially 
as measured
by their internal motions (\cite{jun95}; \cite{jjn96}; \cite{d-pj96}).
Magnetic field structures described below repeat that expectation
with respect to radio emission.

\subsubsection{Magnetic Field Evolution}

Magnetic field evolution is illustrated in Fig. 2 for the ring cloud case.
Fields are modified by compression effects and 
amplified locally through ``field-line stretching'', resulting from
the instabilities and 
associated turbulence inherent to the SNR evolution
and to the shock-cloud interaction.   
Before the cloud-shock collision begins, the ISM magnetic field is
compressed by the supernova blast wave and further amplified by the R-T
instability developing near the contact discontinuity inside the
SNR.  Inside the reverse shock, severe expansion effects reduce the
initial stellar field (simply assumed to be the same as in the ISM) 
to extremely small values.
Since we uniformly aligned
the initial magnetic field along the symmetry axis, the
ISM magnetic field is perpendicular to the SNR blast wave normal
near the equator, but parallel to it at the pole. A
field component perpendicular to a shock normal is increased by strong
shock passage
approximately the same amount as the density compression through the
shock, but the component aligned with a shock normal is unaffected
by the shock. Thus, in our simulations 
blast wave passage increases the magnetic field strength by a 
factor $\sim 4$ near the equator, but not at all near the pole.
At that stage the (weaker) field near the pole is nearly radial
with respect to the SNR center, while the (stronger) field
near the equator is tangential.
That simple description changes in subsequent field evolution, however,
as we shall see.
In passing we comment that, although the initial magnetic pressure in 
the ISM is comparable to the
gas pressure, with $\beta = \frac{p_g}{p_b} \approx 4.5$, the
blast wave is a very strong shock, with Mach numbers $> 200$ even
at the end of the simulations. Thus, the
magnetic pressure just inside the blast wave is dynamically
insignificant, with $\beta > 10^4$.
Even though subsequent field amplification is very substantial,
the magnetic fields remain passive everywhere in these simulations,
with $\beta > 10^2$. We note here and explain further below, however,
that numerical resolution effects and geometry assumptions limit 
field amplification, so it is not meaningful to establish quantitative 
estimates to the strongest fields expected during these interactions.

In addition to greater shock compression of the field near the
equator, the R-T instability associated with the ISM-ejecta contact
discontinuity is also found to amplify more efficiently near the
equator than the pole.  This bias
is explained by the fact that field amplification 
is more effective through field-line stretching along extending R-T fingers
than amplification through
vortical motions generated by the secondary K-H instability (\cite{jns95}).
Magnetic reconnection expels flux from inside vortices (\cite{weiss}; \cite{jgrf97})
limiting field amplification there.
Near the equator, where the field is transverse to the direction of
expansion for the R-T fingers, flux freezing requires that the
field lines be stretched around the R-T fingers, whereas this is
not necessary near the pole, where the field is aligned with the 
fingers.
As the R-T fingers extend, the radial component of the magnetic field 
frozen to them is greatly strengthened (\eg \cite{jun96b}).
At the recent Minneapolis SNR workshop, we proposed this field 
amplification asymmetry as a possible origin for some bilateral radio SNRs
(see \cite{snrmeet}). 
Although we have not directly computed the
polarization properties of the synchrotron emission in
our current SNR models, we note that the stretching of the magnetic
field along R-T fingers leads to enhanced emission from regions with
strong radial fields (see Fig. 4d) even when the field was tangential
to the blast wave at impact. {\it Thus, as noted previously by \cite{jun96b},
the brightest radio emitting regions for a young SNR expanding in a uniform
medium may naturally suggest a preferential radial
field orientation, even when the ambient field is not, and
after initial field compression preferentially enhances the tangential component
rather than the radial component.}
This preferential synchrotron enhancement near the
magnetic equator is an important complication to consider when trying
to ascertain the relative efficiencies of particle acceleration as a
function of angle between the blastwave shock normal and the ambient
magnetic field (Jokipii 1987; Fulbright \& Reynolds 1990). In particular
it would have a very similar outcome to the efficiency effects that
have been posited, so that it would not be possible to establish a
bilateral pattern to particle acceleration efficiency without first
establishing very clearly that the magnetic field {\it inside}
the brightest emitting regions is tangential to the blast wave.
For example, in SN1006 (Reynolds \& Gilmore 1993), Kepler (Matsui \etal 1984)
and Tycho (Dickel \etal 1991) the brightest regions have radial magnetic
fields, so it is likely that postshock turbulence is dominating the surface 
brightness.

Once the cloud is engulfed by the SNR, magnetic field
amplification is enhanced in the vicinity of the cloud, but primarily
outside the cloud itself.
The magnetic field behaviors we see in this region are consistent with those
found in previous MHD studies of plane-shocked clouds (\cite{mmksn94}) and
supersonic clouds (\cite{jrt96}), since the flow patterns in and
around the clouds are similar in all these situations. 
Especially at later times when the cloud begins to be disrupted, we
can see from the magnetic field images in Fig.2 (third column) the
magnetic field near the cloud becomes stronger than in any other
region.  The enhanced field strength around the cloud has two causes.
First, there is some additional field compression by the cloud bow shock,
provided (as in this case) that the initial ISM field has a component
perpendicular to the bow shock normal.
However, as we noted before, in an adiabatic flow the bow shock
compression will never be more than about a factor of two. 
In our spherical cloud simulation (not shown) the initial magnetic field was
mostly aligned with the bow shock normal, so that the field
was not influenced much at all by shock compression.
A much greater effect in both of the cases we computed comes through 
field amplification by the
sheared motions (\ie field stretching), especially coming from the 
combined instabilities developing on the surface of the cloud.   
For example, the strong magnetic field behind the cloud (at larger radii from the
explosion center) 
is particularly noticeable in Fig. 2.
This is a result of strong shear generated along the cloud
perimeter by motion of the relatively faster intercloud shock (the blast wave)
and the slower cloud shock.  An enhanced magnetic field region is
also found at $t = 500$ years outside the Mach disk formed 
following convergence of the intercloud shock onto the cloud
symmetry axis behind the cloud. The field is amplified here
because it is stretched along
the collimated postshock flow (\cite{mmksn94}; \cite{jrt96}).
Comparable features are seen in the spherical cloud case, as well.

In addition to post-cloud magnetic structures, field lines anchored 
in the plasma swept around the
cloud and some of those anchored in the cloud itself are stretched by
the flow around the cloud, enhancing the magnetic field strength in
this region. 
Analogous features were seen in the supersonic cloud simulations of \cite{jrt96}
for situations where the ambient magnetic field was transverse to the motion of
the cloud. Much less field enhancement on the leading edge of the
cloud was seen by them  when the ambient field was aligned to the motion 
of the cloud (or in the analogous shocked-cloud simulations by \cite{mmksn94}
when the field was parallel to the shock normal). 
\cite{jrt96} argued that
evolution of the magnetic field around a moving cloud will
resemble the transverse-field interaction 
provided the angle between the bow-shock normal and the ambient
magnetic field, $\chi$, satisfies the relation $\tan{\chi}~\gsim~\frac{1}{M}$,
where $M$ is the Mach number of the bow wave.
That seems to be consistent with our calculations.

The time history of the mean magnetic energy density inside the
SNRs is shown for six different
simulations in Fig.3.  Three thin lines represent low
resolution simulations (r300x300) for different cloud models, while 
three thick lines show the results for high resolution simulations (r600x600).  
The quantity shown is simply  defined as the integrated magnetic energy
inside the blast wave divided by the associated three dimensional volume.
The early evolution is determined by development of the
R-T instability near the ISM-ejecta contact discontinuity.
Magnetic energy behavior in our cloudless simulations is qualitatively
similar to that seen by \cite{jun96a}, although differences in numerical
resolution prevent us from making quantitative comparisons.
There is a rapid rise in the mean magnetic energy density, peaking 
between t = 50 and 100 years.
\cite{jun96a} demonstrated in their simulations that this peak 
corresponds to the maximum development of
strongly turbulent flow within the SNR. In the absence of a cloud
interaction the amplification and decay of the field then remain very roughly
balanced for the rest of the simulation. As we discuss shortly, the specifics
depend on numerical details.

There are three qualitative
conclusions that we can derive from Fig. 3: 1) There is little
difference in the mean magnetic energy between the case with no cloud 
and the case with a spherical cloud placed on the symmetry axis. 
2) On the other hand, envelopment of the ring cloud placed at a polar 
angle $\theta = \frac{\pi}{4}$
enhances the field amplification inside the SNR significantly compared to the other
two simulations.
The mean field energy after the SNR collides with the ring cloud is
enhanced by about a factor of two compared to the other cases.
Most of this rise takes place over roughly 150 years, during the interval that the
cloud is being enveloped by the blast wave and colliding with the irregular
ejecta-ISM contact discontinuity.
There are a couple of reasons that the mean magnetic field is more strongly
enhanced during the collision with the ring cloud.
Both are geometrical in origin.
First, since the initial cloud radius, $r_{co} = \frac{1}{4} r_{cc}$,
where $r_{cc}$ is the central position of the ring cloud intersection
with the computational
plane, the ring cloud occupies a much greater three-dimensional volume 
($ 2\pi^2 r^2_{co} r_{cc}$) than the spherical cloud ($\frac{4}{3}\pi r^3_{co}$).
Therefore, the enhanced
magnetic field region in the ring cloud simulations is larger than in the
spherical cloud simulations. That is partially offset by the fact that
the cloud interior, which is also larger for the ring cloud, has very little
field amplification. Second and more important
physically, the amplification of magnetic
field is much more efficient during the interaction with the ring cloud,
because the magnetic field has a component 
transverse to the SNR blast propagation.  
This point was emphasized earlier, in the discussion about the magnetic field
evolution around the clouds.
3) By comparing the equivalent simulations done with different numerical
resolutions, we see that the magnetic energy enhancement is greater when the numerical
resolution is greater. This behavior is a well-known property of ``ideal'' MHD
simulations of disordered flows (\eg \cite{mmksn94}; \cite{jns95}; \cite{jgrf97}).
It results from the fact that in such simulations
magnetic and gasdynamical structures are limited on the
smallest scales by numerical diffusion. Higher resolution allows
smaller structures to be captured. In complex flows these 
are often places where fields should be locally stretched and
amplified or where magnetic reconnection should take place when local current
sheets form (\ie when $\vec\nabla \times \vec B$ is large). Thus, 
we can only asymptotically approach capturing these elements.
Finally, we note that since our initial magnetic field has an
azimuthal symmetry, no dynamo can develop to enhance large-scale 
fields permanently. These issues will be revisited in the subsequent
discussion of radio emission computed for the models.

\subsubsection{Relativistic Electrons and Radio Emission}

The relativistic particle distribution computed as described in
\S 2 and the magnetic field structure just described enable us to
compute the radio synchrotron emissivity self-consistently within our model SNRs.
As far as we are aware these are the first multi-dimensional SNR
simulations from which that could be done self-consistently, since
that calculation requires the properties of both the particles and the
fields.
In a very recent publication \cite{sturn97} have computed for some
1-D SNR models radio to $\gamma$-ray integrated spectra including
a simple model for DSA. While their work does include a wide variety
of nonthermal processes and is, therefore, quite valuable, since it
is 1-D it cannot examine at all such issues as the important influence
of the magnetic field and particle distributions within the SNR.
Our results are in qualitative agreement with theirs to the extent
that they overlap, however.

The nonthermal radio emissivity as
a function of frequency can be written in terms of the
parameters we defined in \S 2 for the distribution function, $f(\vec r, p)$, 
as (see \eg Jones \etal 1974) as
\begin{equation}
j_\nu = 4\pi j_{\alpha o} \frac{e^2}{c}\frac{1}{(mc)^{2\alpha}} f_i p_i^{2\alpha+3} ~
\nu_{B_\perp}^{\alpha+1} \nu^{-\alpha},
\label{emiss}
\end{equation}
where $\alpha = \frac{q-3}{2}$ is the usual synchrotron
spectral index, $j_{\alpha o} \sim 1$ is a
dimensionless function of $\alpha$ tabulated in Jones \etal 1974, $\nu_{B_\perp}
= \frac{1}{2\pi}\frac{e B_\perp}{mc}$ is the
electron cyclotron frequency for the magnetic field projected
onto the plane of the sky, and the other physical constants take
their usual meanings.
Equation [\ref{emiss}] is valid in the frequency range $\nu_i\lsim\nu\lsim\nu_{i+1}$,
where $\nu_i = (p_i/mc)^2 \nu_{B_\perp}$, etc, assuming $p_i/mc >> 1$.
Recall that $f_i$ refers to the momentum bin
boundary $p_i$, as defined in association with equation [\ref{dndt}].
Expressed in terms of the momentum index, $q$, and the magnetic
field, $B$, the synchrotron emissivity, $j_\nu$ is
\begin{equation}
j_\nu = C_q p_i^q f_i {B_\perp}^{(q-1)/2} \nu^{-(q-3)/2},
\end{equation}
where we have lumped together into $C_q$ all the physical constants 
that are independent of our simulations.
Note, further, that the remaining derived physical variables, $B,~q,~{\rm and}~f_i$
are all computed explicitly in our simulations.
For the single power-law electron distribution utilized in the
present paper, $p_i^{q} f_i~\approx  n_1~(\frac{q-3}{4\pi}) p_1^{q-3}$,
provided $p_i >> p_1$ ($i > 1$) and $q > 3$,
where $n_1$, is the number density of
relativistic electrons for $p_1\le p \le p_2$. Both $n_1$ and $q$ may
vary with location, of course. Under the conditions that apply here
$n_1$ is also approximately equal to the total
number density of relativistic electrons; \ie those above 1 MeV. 
In addition our magnetic field
is two-dimensional, so as long as we restrict ourselves to equatorial
views of the computational plane the magnetic field vector is entirely in the plane of the sky.
As we shall see below, $q$ varies relatively little in the inter-shock
region of the model SNRs.
Thus, we can characterize the synchrotron emissivity spatial variation 
in terms of a function, $j_o$, defined as 
\begin{equation}
j_o \approx  C_o n_1 B^{(q-1)/2},
\label{simpj}
\end{equation}
where we have subsumed the frequency dependencies into a new factor, $C_o$,
which is almost constant over regions of interest.
This emissivity is the same as that employed by \cite{jon93} in their
examination of particle acceleration associated with a plane-shocked
cloud and by \cite{jkt94} in a similar study for supersonic clouds.

In equation [\ref{simpj}] $j_o$ scales directly with
the electron density in the lowest momentum bin, $n_1$, whereas, 
since typically $q \approx 4$, it scales with magnetic field
as $B^{\frac{3}{2}}$.
The simple ``density'' dependence comes about because $n_1$ follows
electrons in a fixed momentum range. Our computational method 
of computing $n_1$
automatically compensates for adiabatic compression effects on the 
energetic particles.
This representation is, therefore, different from
the heuristic emissivity model introduced by Clarke \etal  (1989), which assumes
$j\propto P_g B^{\frac{3}{2}}$, where $P_g \propto \rho^{\frac{5}{3}}$ 
is the thermal  gas pressure.
The use of a pressure term in that model rather than a density
represents an implicit effort to follow adiabatic effects on the particles.
According to our equation [\ref{simpj}] the radio emission is more sensitive to
the magnetic field strength than the relativistic electron number
density, but neither contribution 
should be neglected in interpreting the origin of radio emission.

Fig.4 and Fig.5 show magnetic field strength, relativistic electron number
density, electron power law index, and resulting radio emissivity at t= 200 years
and t = 500 years, respectively for the ring cloud simulation. 
These also correspond to the earliest and latest times shown in Fig. 2.
The cloud in Fig. 4 is only partially enveloped by the
SNR blast wave and it has not yet penetrated the unstable 
ISM-ejecta contact discontinuity. Its influence on the magnetic
energy in the SNR is still developing.
By contrast, the cloud in Fig. 5 is fully enveloped by the SNR blast
and the ISM-contact discontinuity is highly deformed by the cloud. 
While still intact, the cloud is showing clear signs of disruptive instabilities
and has produced some strongly enhanced magnetic field
structures. 
At the earlier time there are two distinct regions of relativistic
electron concentration visible (Fig.4b). One is the region between
the reverse shock and the unstable ISM-ejecta contact discontinuity.
As pointed out in the discussion of analogous one-dimensional
simulations (\S 4.1) the reverse shock is a highly effective site of
particle acceleration in these simulations. This is very important, 
in fact, because of a new feature that shows up in these two-dimensional
simulations; namely, that some of the reverse-shock accelerated
particles are embedded in the region
where magnetic fields are strongly amplified by R-T instabilities. 
Thus, they contribute significantly to the radio emission from the ``turbulent''
contact discontinuity. 
That is evident by comparing Fig.4b and Fig.4d.
Examination of the passive mass fraction labeling stellar
ejecta shows that electrons from both the forward and reverse shocks are
in these bright structures, but the details are numerically determined.

From Fig.4b we can see in addition that at $t = 200$ years, electron injection is 
also strong inside the cloud.
This behavior was previously seen for cloud shocks in the
simulations reported by \cite{jon92} and \cite{jkt94}.
The reasons are the same as they were for the SNR reverse shock; \ie
the cloud shock is strong and the flux of thermal electrons that
are available for injection is quite large. This leads the portions
of the cloud body that have been shocked to be among the regions
with the highest density of CR electrons at this time.
Since the magnetic field in some of those same cloud portions is
strong due to shock compression and, especially, field shear along
the edge of the cloud, the leading portions of the cloud also
are radio bright. The integrated radio synchrotron spectral index at this time
would be very close to $\alpha = 0.5$, since virtually the entire
electron population in the region enclosed by the shocks mentioned
and containing enhanced magnetic field
has a momentum index, $q = 4.0$. That is apparent in Fig.4c, where
the black color inside the computational grid corresponds to $q = 4.0$.

By $t = 500$ years the situation has become more complex. As Fig. 5
shows clearly, the ISM-ejecta
contact discontinuity has begun to interact with the cloud, the
cloud shock has exited, the tail-shock structure behind the cloud
has formed and instabilities have begun to disrupt the cloud. This
combination leads to relatively pronounced synchrotron emissivity along the
perimeter of the cloud and in its wake. That is very similar
to the pattern found for a plane-shocked cloud by \cite{jon92} 
when the incident magnetic field was transverse to the shock normal and
shock-injected electrons dominated the CR electron population (see
their Fig.9c). By this time the brightest regions of radio 
emission in our simulation are dominated by the growth of instabilities and
stretching of magnetic fields.
The rapid rise in magnetic energy seen here in Fig.3 for 
the ring cloud simulation and the near factor of two excess
radio emission from the SNR in this case visible in Fig.6d
by $t = 350$ years represent the same phenomenon.
\cite{jkt94} showed for similar
reasons that the radio synchrotron emission from a 
supersonic cloud grew rapidly and was concentrated along the
cloud perimeter as instabilities began to disrupt that cloud.

The electron spatial distribution in Fig.5b resembles the total
gas density distribution (compare Fig.2), but that
is rather different from the fine structure of the synchrotron emissivity. 
On the whole fine structured synchrotron emission regions seem to
reflect enhanced magnetic field structures, particularly tips of R-T
fingers while smoother, lower surface brightness emission seems more 
representative of the gas density distribution.
The integrated spectral index remains close to $\alpha = 0.5$ to the
end of our simulations.

However, an interesting development between $t = 200$ years and $t = 500$ years is 
that at the later time some
radio bright regions have momentum indices 
$q \sim 4.2~-~4.4$ and, hence, the radio spectrum there has an index
$\alpha \sim  0.6 ~ - ~ 0.7$ (compare Fig.5c and Fig.5d). That results from
a combination of two effects along the cloud perimeter. First, while
a very strong cloud shock penetrates from front to back in the cloud
interior, a much weaker shock moves into the cloud from the
sides, driven by the high pressure of the inter-shock region. That is
apparent in the density images of Fig.2 at $t = 200$ years and $t = 300$ years.
Compression through these shocks is often only a factor $\sim 3$.
Electrons injected and accelerated at those
weaker shocks consequently have relatively steeper spectra than
those in the strong shocks. Later, those same
cloud boundary regions are mixed by instabilities with strongly
shocked gas and local magnetic fields are amplified. With our simple 
single power law model for the electron
momentum distribution, mixing leads simply to an intermediate
slope for the electron population. A more refined model would lead to
a concave momentum distribution, tending from the values quoted above
at low frequencies to $\alpha \sim 0.5$ at very high frequencies.

We note here that while observations of galactic SNRs show a
distribution of radio spectral indices roughly centered on $\alpha = 0.5$,
there is a fairly wide spread, with some tendency of
younger remnants to have steeper spectra (\eg \cite{green}; \cite{snrmeet}).
While observational uncertainties contribute to the measured
distribution (D. Green, cited in \cite{snrmeet}) it is unlikely
that young SNRs uniformly have radio indices $\alpha = 0.5$.
On the face of it that seems inconsistent with our results, which
give $\alpha$ very close to $0.5$ as long as the SNR
blast is strong. There are at least two effects short of introducing
other acceleration physics that could ameliorate
this contradiction and that should be examined in future investigations
of this kind. 
First, our simulations treat only CR electrons, but energetic
protons are certain to be accelerated at the same time. It is well known
that leakage of those protons upstream can produce sufficient
pressure to adiabatically compress
the upstream gas (\eg \cite{drury83}). While this increases the
total compression through the shock, it also tends to reduce the
compression through the dissipative ``subshock''. For particles
of relatively low energy, such as the electrons responsible for
radio emission in SNRs that can lead to a steeper spectrum than the
value associated with a gasdynamic shock (\eg \cite{rey92}). 
Some simulations of SNR blast waves including dynamical influences of CR protons
suggest that these effects may be significant for relatively
young SNRs at dynamical ages analogous to those in
these simulations (\eg \cite{jon93}; \cite{ber95}; \cite{ber97}). 

A second possibility is closer to the focus of the present paper. We noted
just above that some of the synchrotron emission from the perimeter of the cloud in
our simulation should have a relatively steep spectrum because of the presence
there of weak shocks. The volume was, however, too small in the
present simulations to have a measurable
influence on the total SNR spectrum. On the other hand in a clumpy ISM
there would be possibly many such volumes. 
Similar
to the situation for small supersonic clumps as described by \cite{and94}
the net spectrum of each clump should be steeper than $\alpha = 0.5$
even if the cloud motion is highly supersonic through the SNR material.
In addition, there will be interactions between bow shocks of individual
clumps, and the subsequent blending of those structures via the
turbulence generated within the SNR by the presence of the clumps 
(\cite{jun95}; \cite{jjn96}). Those effects could produce a measurable
influence on the net spectral index of an SNR. Also, as perhaps detected in
Cas A (\cite{and96}), there should then be some measure of spatial variation
of spectral index within the remnant, the degree of variation dependent on the 
degree to which these effects are important and the amount of mixing within the remnant.

The time history of the mean CR electron energy density is shown  for the
three high resolution simulations we have done in Fig.6a.
The relativistic energy density is proportional to the number density
of relativistic electrons.  Therefore, the relativistic electron energy
density is a sensitive function of thermal electron injection.  Since
the injection rate at a shock is proportional to the incoming flux of thermal
electron, \ie $n_e v_s$, where $n_e$ is the thermal electron number
density and $v_s$ is the shock velocity, we can see also that the
relativistic electron energy density  immediately behind a shock
is proportional to the upstream
gas density and the shock velocity.  By looking at the cloud-free simulation
in Fig.6a, we can see that the relativistic electron energy
density generally decreases with time because of the decreasing shock
velocity.  There is a kind of plateau in the electron
energy  density for this model while the reverse shock is passing through 
the relatively dense shell of ejecta present during the
early self-similar phase of evolution. That ends abruptly around $t=220$
years when the gas density interior to the reverse shock drops suddenly.
The self-similar stage thusly  ends as the reverse shock reaches the inner
core region of the supernova gas (see \cite{jun96a} for the
details of the global dynamics of SNR).  Simulations  including the 
cloud  interaction show a sudden enhancement of relativistic electron energy beginning
about $t= 100$ years.  This result is, of course, due to the cloud
encounter and associated increased injection of CR electrons.  The
energy density in the simulation with the ring cloud is larger than the
case with the spherical cloud because of the larger filling factor of the
ring cloud mentioned before.   
The total relativistic electron energy in Fig.6b increases until the
end of all the SNR simulations
because injection continues and because we measure this energy within a 
fixed range of energies.
There is a noticeable inflection in this curve for the cloud-free
and spherical cloud cases as the SNR
evolves out of the self-similar stage.   The enhanced
total relativistic electron energy in the presence of the ring cloud is
also apparent.   The simulation result with a ring cloud shows a lower
relativistic electron energy than other cases after t=420 years.  
This is likely due to the retarded propagation of the forward shock after
the interaction with the cloud reducing the total mass swept
by the shock (see Fig.3).

The time history of the radio emission is shown in Fig.6c and 6d.  The
volume-averaged radio emissivity is plotted in Fig.6c.  The
enhancement of the radio emission by the cloud interaction is once
again much more evident in the ring-cloud case and more
pronounced than the relativistic electron energy enhancement
because magnetic energy increases, which are greater during the
cloud encounter, are more important to the radio emission. 
In general, the radio luminosity
is an increasing function of time throughout the interval we have
studied.

The radio luminosity in Fig.6d is normalized by the final value for
the high resolution ring-cloud simulation. Using the full version of
the emissivity given in equation [\ref{emiss}] this value  at a 
frequency of 1 GHz is 
$L_{\nu} = 7.4 \times 10^{19}$ erg-cm$^{-2}$sec$^{-1}$Hz$^{-1}$. 
It is interesting to compare that figure with the actual radio
luminosity at 1 GHz for Kepler's SNR, which is of comparable age and is
usually presumed to be formed from a Type Ia SN (see \cite{ban87} for a
discussion of the issue).
Kepler is measured to have a spectral flux at 1 GHz of 19 Jy
and an angular diameter of 3 arcmin (\cite{green}). The distance is
very uncertain, but values in the 3-5 kpc range are commonly estimated
(\eg \cite{blv91}). That gives a physical radius $\sim 1-2$ pc
and a spectral luminosity $\gsim 10^{23}$ 
erg-cm$^{-2}$sec$^{-1}$Hz$^{-1}$, the latter being about 3 orders of magnitude greater
than the simulated radio luminosity. A similar comparison would result if
we had chosen Tycho's SNR  or SN1006 as other possible Type Ia examples, since
their luminosity, age and size are similar to Kepler (\cite{green}).

There are two possible contributors to the luminosity difference; namely, 
the number of radiating electrons and the strength of the magnetic field
in each remnant. We have injected into the CR population $10^{-4}$ of the electrons available
in about 7 M\sun ($\sim 10^{54}$ electrons).  About 0.1\% of the
injected electrons are relativistic. 
Independent of what process is responsible for electron acceleration
it is extremely unlikely that the number of electrons available from
the mass included in our simulated SNR could
be increased by more than about an order of magnitude. Even if we
suppose that
Kepler was a Type Ib SN (\cite{ban87}) and perhaps an order
of magnitude more massive in ejection, it seems unlikely that
the major cause of the apparent deficiency in the simulated radio
luminosity is the number of electrons available.

It is much more likely that the characteristic magnetic fields in
such SNRs as Kepler and Tycho are larger than in the simulated 
remnants. 
We can use equation [\ref{emiss}] to estimate a ``characteristic''
field needed to produce a 1 GHz luminosity $10^{23}$ erg-cm$^{-2}$sec$^{-1}$Hz$^{-1}$
from the electron population available in our simulated SNR.
By substituting the total electron number for the volume number into
equation [\ref{emiss}], we find, independent of the radius of the remnant, 
that characteristic field to be $B_o \approx 4\times 10^{-4}$ Gauss. This
corresponds to $\beta \sim 10$ in much of the model SNR shell; 
that is,  it is moderately close to equipartition
between the magnetic field and the thermal energy in the plasma.
Although magnetic field strengths in the simulations reach
values as large as $10^{-4}$ Gauss, corresponding to $\beta \sim 10^2$,
the magnetic field is much weaker in most of
the volume occupied by the relativistic electrons. Recall that relativistic
electrons
are distributed much like the thermal plasma.

What field ``characterizes'' the synchrotron emission in our simulated
SNR? We can again use equation \ref{emiss} 
applied to the simulated synchrotron luminosity. That gives
$B_s \approx 3\times 10^{-6}$ Gauss,
which is coincidentally close to the assumed circumstellar field. This result
emphasizes that the filling factor for strong fields in the
simulations is small, and that there are regions inside the model SNRs
where the fields are quite weak. We touched in \S 4.3.2 on the issue of
magnetic field evolution in numerical simulations of this kind.
Finite numerical resolution certainly limits the growth of fields, as mentioned
there. That, however, is probably not the principal effect in this
case, since the simulated radio luminosities found using the
lower resolution (r300x300) results differ by only about 10\% from
the higher resolution results. That lack of sensitivity comes from the
small filling factor for the stronger field regions in either case;
that is from the fact in the simulations that most of the
synchrotron luminosity actually comes from regions of relatively
low emissivity. From this is it clear that the issue is not just
what maximum strength the fields might achieve, but on the magnetic
``intermittency'' (to use the appropriate term from MHD turbulence
literature).
If the external medium were generally clumpy, the high emissivity
filling factor would certainly be larger (\cite{jun95}; \cite{jjn96}).
However, the assumption of azimuthal symmetry is probably the dominant
limitation. That eliminates any kind of ``dynamo'' action that might
amplify large-scale fields within the SNR in response to turbulent 
or convective motions. Thus, fully three-dimensional simulations
of high resolution will be required in order to model quantitatively
synchrotron emission from SNRs. That fact does not, however, diminish
the value of two-dimensional simulations as an initial guide to
establish the qualitative role of nonspherical dynamics in 
understanding radio emission from SNRs.

We add one final brief comment on possible implications for nonthermal
X-ray emission from remnants such as we have simulated. As discussed in
\cite{snrmeet} and references therein, there are two possible sources
of nonthermal X-rays. Relativistic electrons with energies $E \gsim 10^{14}$ eV
might produce synchrotron emission in the X-ray band. We have not directly
modeled acceleration to such high energies. However, since our nonthermal
electron populations are predominantly accelerated in very strong shocks,
we would expect the X-ray synchrotron emissivity to resemble qualitatively the 
radio emissivity that we have computed. The second possibility is
nonthermal bremsstrahlung from electrons only somewhat super-thermal
in the non-Maxwellian tail resulting from DSA, for example. Those
populations may also influence X-ray line emission, mimicking the
presence of a hotter thermal plasma. We have crudely examined the
comparative distributions of thermal and nonthermal bremsstrahlung
emissivities in the models by comparing the spatial distributions
of $j_\nu (therm) \sim \rho^{\frac{5}{2}} P^{-\frac{1}{2}} \propto n^2_e T^{-\frac{1}{2}}$
and $j_\nu (nontherm) \sim \rho n_1$. 
To the lowest approximation they are the same, since the nonthermal 
electron population mirrors the thermal plasma density. There is, however,
some relative enhancement in $j_\nu (nontherm)$ in the regions
associated with the RT fingers, since the nonthermal electron population
is relatively large there (partly due to acceleration in the
reverse shock). Thus,
the nonthermal bremsstrahlung distribution would resemble a hybrid between the
thermal emission and the synchrotron emission. It would be inappropriate
to estimate the strengths of these nonthermal components, since that
would depend on the actual electron injection efficiency, $\epsilon$, and
the actual magnetic field distribution. Neither can be reliably
determined in simulations of this kind, as argued above.

\section{Conclusions}

We have carried out two-dimensional MHD simulations of the interaction
between a young SNR and an interstellar cloud
comparable in size to the SNR at the time of impact. This produces
a more complex interaction than the previously studied collision 
between a plane shock and a cloud, primarily because the SNR
blast wave is soon followed in our simulations by the shell of stellar ejecta.
From the perspective of what happens to the cloud, however, it is
qualitatively similar to a collision with a plane shock. The cloud is
crushed by the shock passage and then destroyed by R-T and K-H
instabilities. At the same time penetration of the cloud into the SNR 
interior substantially
disturbs flows within the SNR, especially as it encounters the contact
discontinuity between the ISM and the stellar ejecta.
We have developed a simplified but self-consistent multi-dimensional
model for the associated synchrotron radio emission 
that includes the evolution of magnetic
fields and the relativistic cosmic-ray electron population 
accelerated by shocks in the SNR.  The magnetic fields are evolved 
through quasi-ideal MHD while the transport of relativistic
electrons is treated through a simplified formulation based on
the standard momentum-dependent diffusion advection equation.
Diffusive acceleration at shocks is included in the approximation
that the electron distribution comes to equilibrium in
the momentum range of interest with a power law distribution
determined by the compression through the shock. 
Fresh injection of cosmic-ray electrons at shocks is included
through a standard model that injects a small, fixed fraction of
the thermal electron flux into the nonthermal population.
CR electrons outside the shock itself are diffusion-free and can
be treated by an advection equation including adiabatic losses.

From these calculations we identify the following primary results:\\
(1) Independent of the cloud interaction, the SNR reverse shock can be an efficient
site for particle acceleration, because it is stronger
than the SNR forward shock, and because during much of its existence
the mass flux (and hence the thermal electron flux) is
greater than through the forward shock.  Thus, at these times
it should not be ignored as a source of energetic particles.
Its possible role as a source of radio emission would depend on there
being some mechanism to maintain at least a moderate magnetic
field in the ejecta, however. The relativistic electron
spatial distribution qualitatively resembles the gas density distribution 
within the SNR inter-shock region. \\
(2) The flow inside the SNR becomes highly turbulent once it encounters
the cloud. This is due to
interactions of many different features: a reflected cloud bow shock,
a transmitted cloud shock, eventual cloud disruption, the Rayleigh-Taylor 
instability in the interaction region, as well as distortions of the  
supernova blast wave and the reverse shock.
The turbulent flow is an efficient means to amplify the magnetic field,
while the relativistic electrons are accelerated through
many different shocks.  The reverse shock and forward shock are
severely distorted by the cloud encounter. In a cloudy
external medium the interior of such a young remnant should be
highly disturbed.\\ 
(3) An initially uniform magnetic field is much more strongly
amplified 
near the magnetic equator compared to the region near the magnetic
pole. 
That results primarily from the different interactions between magnetic
fields tangential or aligned with the expansion directions of Rayleigh-Taylor
fingers within the SNR.
Amplification of magnetic field associated with R-T finger growth will lead to
a magnetic field that is radial within the SNR, however.
The biased amplification of magnetic field near the magnetic equator is also 
found in the presence of the cloud. If the cloud is placed near the the magnetic
pole of the SNR there is relatively little augmentation to the
magnetic field in the SNR that results from the cloud encounter,
whereas interaction with a cloud when the prevailing field crosses
the cloud significantly enhances magnetic field development in the
SNR during its envelopment. This means that radio synchrotron
emission could be much stronger in equatorial regions, even
leading to a ``barrel'' morphology. Since that goes in the
same direction as some predictions of field-geometry-dependent
particle acceleration efficiency (e.g., Fulbright \& Reynolds 1990)
it offers some strong need for caution in interpreting 
measurements of asymmetries in synchrotron brightness.
In particular any effort to argue for asymmetries in particle
acceleration must first determine clearly that the magnetic field
in the brightest emitting region is tangential to the blast wave. Otherwise,
it is likely that magnetic field amplification asymmetries are
dominant.\\
(4) The filamentary and
turbulent structures of the brightest radio emission are found to correlate well with
the magnetic field in general, while the smoother, low surface brightness
emission reflects the generally smoother spatial distribution of the
electron population, which mirrors the gas density. 
The simulated radio luminosities are significantly
smaller than some young Type Ia SNRs, such as Tycho, Kepler and SN1006. 
This probably comes from numerical limitations, particularly the assumed
2D, axial symmetry that restricts the growth of a pervasive, strong
magnetic field.\\
(5) The cosmic-ray electron momentum index is close to 4.0 in most
of inter-shock region, 
because both the reverse shock and the blast wave are strong.  
That corresponds to a synchrotron index $\alpha = 0.5$.
This result is fully expected.
However, it is important to notice that
steeper power law indices ($\sim 4.2~-~4.4$ corresponding to
synchrotron indices $\alpha \sim 0.6~-~0.7$) are found near the
disrupting cloud.  The steeper power law index results from particle acceleration
at the weak shock in the sides of the cloud and by mixing
between the associated gas and that processed by the stronger
primary shocks.
We point out that such sources of steeper spectra might be a
ubiquitous feature of SNRs accelerating electrons by the
diffusive shock process, especially if they are propagating into
highly inhomogeneous media. The simple 2 momentum bin
model for electron propagation that we used for these simulations
constrains the electron distribution to a pure power. In a
multi-bin extension of this
treatment, regions of mixed populations would actually have a
concave spectral character, since the highest energy electrons would
be dominated by stronger shock sources. On the other hand, it is not clear how
easy that signature would be to detect in SNRs, since the large
range in magnetic field strengths found in these same regions
will spread out the characteristic emitted frequencies for electrons
of a given momentum. Nonetheless, it may be worthwhile to examine
in detail the radio spectral distributions in young SNR interiors
to see what patterns associating kinematics, smoothness of surface
brightness and spectrum might be detectable. Those, in return may
help to identify the dynamical features most responsible for
the acceleration of a given electron population.
Our simulations make it apparent that the spectrum may not be a simple
indicator of the compression through the primary shocks in the SNR.
Thus, it will be important to examine other indicators of
internal source dynamics, such as Doppler motions of thermal
plasma before interpreting spatial variations in the radio
spectral index.\\
(6) The relativistic electron population in the SNR is 
significantly enhanced by the cloud
interaction in our simulations, because the cosmic-ray
injection from thermal electrons is high at the strong shock inside 
the cloud.
The enhanced relativistic energy and magnetic field due to the cloud
encounter result in a significantly higher radio luminosity.

\acknowledgments

The work reported here is supported by NSF grants AST-9318959 and
AST-9619438 and by the Minnesota Supercomputing Institute. We are grateful to
L. Rudnick for many discussions and useful comments on the manuscript.

\clearpage

\figcaption[fig1.ps]{ One-dimensional numerical simulation of a young
Type Ia supernova remnant expanding into a uniform ISM. 
Panel a shows the electron number density in  the energy
range 1~MeV$ < E < 1~$GeV based on a simplified model for
diffusive shock acceleration, including cosmic-ray injection from the
thermal population as discussed in the text.  Plots
represent the electron number density profile as a function of
radius at t=100 years, t=300 years, and t=500 years, respectively.  R.S.,
C.D., and F.S. stand for reverse shock, contact discontinuity, and
forward shock, respectively.   Panel c shows the electron number
density in the energy range of 1~GeV $<$ E for the same case as Fig.1a.
Panels b and d show the same simulation as panels a and c except that
the cosmic-ray injection process is omitted.  The dotted lines in panels b and d
represent the electron number density including adiabatic
effects, but excluding diffusive shock acceleration.
The solid lines show the results including 
shock acceleration.  Panel e shows the power law index of the
electron momentum
distribution.  Background electrons are given an index
4.7. Electrons between the reverse and forward shocks  have
been left with a momentum spectral index very close to 4.0
by diffusive shock acceleration in the two shocks.
comparison, panel f shows the gas density radial
distributions at the same times.  \label{fig1.ps}}

\figcaption[fig2.ps]{Two-dimensional ($r600\times 600$)
simulation of a young Type Ia supernova
remnant interacting with a ring-shaped cloud.  Panel columns from 
left to right respectively
represent gas density, mass fraction of cloud, and magnetic intensity.
From top to bottom, each column follows
the time evolution with images at t=200, 300, 400, and 500 years.  
White tone
corresponds to the highest value in each image.  Structures visible from the
inside outwards include the reverse shock,
contact discontinuity (with R-T fingers), shocked cloud, and supernova
forward shock. The computational boundary is distinctly visible in 
the density and magnetic intensity images
as a circle tangent to the outer box boundary, 
since the simulation is carried out in
spherical coordinate and remapped to cylindrical coordinates for
display.   In the
magnetic field image the reverse shock cannot be seen because
expansion of the original stellar field has severely reduced
the field strength. \label{fig2.ps}}

\figcaption[fig3.ps]{Mean magnetic energy density evolution for various
cloud models and numerical resolutions.  The legend for each line is
shown in the plot. \label{fig3.ps}}

\figcaption[fig4.ps]{Greyscale images of magnetic field strength (a),
relativistic electron number density (b), electron momentum power law index
(c), and radio emissivity (d) at t=200 years for the $r600\times 600$
ring cloud simulation.  White
tone corresponds to the highest value in each image.  In panel c, white 
corresponds to the index 4.7 and black corresponds to
the index 4.0. The region outside the computational grid is also
given a black tone in each image.  \label{fig4.ps}}

\figcaption[fig5.ps]{Same greyscale images as Fig.4 representing
t=500 years. \label{fig5.ps}}

\figcaption[fig6.ps]{History of the spatial mean relativistic electron energy
density (a), total relativistic electron energy (b), spatial average
of radio emissivity (c), and radio luminosity (d), respectively.  The
solid line, dotted line, and dot-dashed line represent the evolution
without the cloud, with a ring shape cloud, and with a spherical
cloud, respectively. Each plot is normalized by the largest
value on any curve in the plot.
\label{fig6.ps}}

\end{document}